\documentclass[conference]{IEEEtran}
\IEEEoverridecommandlockouts
\usepackage{cite}
\usepackage{amsmath,amssymb,amsfonts}
\usepackage{algorithmic}
\usepackage{graphicx}
\usepackage{textcomp}
\usepackage{xcolor}
\IEEEoverridecommandlockouts
\usepackage{cite}
\usepackage{color}
\usepackage{amsmath,amssymb,amsfonts}
\usepackage{algorithmic}
\usepackage{graphicx}
\usepackage{textcomp}
\usepackage{xcolor}
\usepackage{multirow}
\usepackage{booktabs}
\usepackage{algorithm}
\usepackage{color}

\def\BibTeX{{\rm B\kern-.05em{\sc i\kern-.025em b}\kern-.08em
		T\kern-.1667em\lower.7ex\hbox{E}\kern-.125emX}}
\begin{document}
	
	\title{On The Design of a Light-weight FPGA Programming Framework for Graph Applications
	}
	
	\author{
		\and
		\IEEEauthorblockN{Jing Wang}
		\IEEEauthorblockA{Department of Computer\\ Science and Engineering\\
			Shanghai Jiao Tong University\\
			Email: jing618@sjtu.edu.cn}
		\and
		\IEEEauthorblockN{Jinyang Guo}
		\IEEEauthorblockA{Department of Computer\\ Science and Engineering\\
			Shanghai Jiao Tong University\\
			Email: lazarus@sjtu.edu.cn}
		\and
		\IEEEauthorblockN{Chao Li}
		\IEEEauthorblockA{Sustainable Architectures \\and Infrastructure Laboratory\\
			Shanghai Jiao Tong University\\
			Email: lichao@cs.sjtu.edu.cn}
	}

	\maketitle
	
	\begin{abstract}
		
		FPGA accelerators designed for graph processing are gaining popularity. Domain Specific Language (DSL) frameworks for graph processing can reduce the programming complexity and development cost of algorithm design. However, accelerator-specific development requires certain technical expertise and significant effort to devise, implement, and validate the system. For most algorithm designers, the expensive cost for hardware programming experience makes FPGA accelerators either unavailable or uneconomic. Although general-purpose High-Level Synthesis (HLS) tools help to map high-level language to Hardware Description Languages (HDLs), the generated code is often inefficient and lengthy compared with the highly-optimized graph accelerators. One cannot make full use of an FPGA accelerator's capacity with low development cost. 
		
		To easily program graph algorithms while keeping performance degradation acceptable, we propose a graph programming system named JGraph, which contains two main parts: 1) a DSL for graph atomic operations with a graph library for high-level abstractions including user-defined functions with parameters, 2) a light-weight HLS translator to generate high-performance HDL code, cooperating with a communication manager and a runtime scheduler. To the best of our knowledge, our work is the first graph programming system with DSL and translator on the FPGA platform. Our system can generate up to 300 MTEPS BFS traversal within tens of seconds.
		
	\end{abstract}
	

	
	\section{Introduction}
	
	Graph is the basis of many complex algorithms. As shown in Table \ref{tab1}, today many important applications such as social networks and recommendation systems rely on graph analysis of vertices and edges. Graph processing can be a challenging task. For example, power-law graphs with irregular degree often aggravate random memory access, which results in poor locality when performing traversal\cite{gp-memory}. 
	 
	Heterogeneous architectures are often adopted because they can achieve much better performance in large-scale graph processing than traditional CPU-based architectures\cite{wpy,frog,gunrock}. In particular, FPGA accelerators\cite{graphgen,graphicionado,grafboost,gps,gravf,gravf-m,hust1jin,hust2yao,hmc1zhangli,hmc2zhangli,hmc3zhangli,zhou1high,zhou3hitgraph,foregraph,fpgp} have attracted many attentions in the graph processing field, due to high performance and efficiency. FPGAs are also famous for reconfigurable hardware design style with flexible interconnection of logic units and fine-grained parallelism. 
	
	\begin{table}[t]
		\caption{Graph processing applications and algorithms}
		\label{tab1}
		\begin{tabular}{c|c|c|c}
			\hline
			\multicolumn{1}{l|}{\textbf{Applications}} & \multicolumn{1}{l|}{\textbf{Vertices}} & \multicolumn{1}{l|}{\textbf{Edges}} & \multicolumn{1}{l}{\textbf{Algorithms}} \\ \hline
			Social network                             & individual                             & friendship                          & PR/BFS/DFS                              \\ \hline
			Electronic Commerce                        & Customer                               & transaction                         & BC/TC/SSSP                              \\ \hline
			telecommunication                          & phone                                  & Conversation                        & SSSP/MM                                 \\ \hline
			Supply chain                               & Supplier                               & Channel                             & DFS/BFS/SSSP                            \\ \hline
		\end{tabular}
	\end{table}
	
	\begin{figure}[h]
		\centering
		\includegraphics[width=\linewidth]{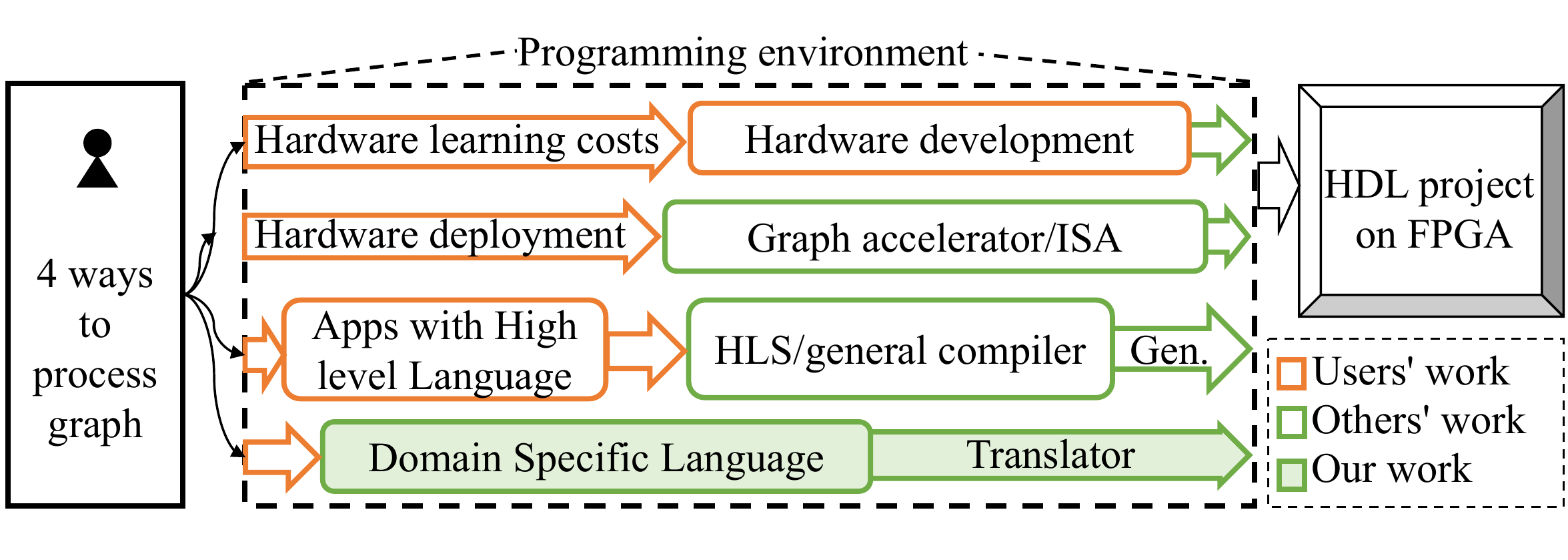}
		\caption{Graph basic programming environment comparation on FPGA}
		\label{fig-apps}
		
	\end{figure}
	
	Unfortunately, there is a considerable gap between algorithm developers and highly-optimized accelerators, which is called the "programming wall"\cite{wall}. The cost for hardware programming isolates FPGAs performance gaining. There are two main reasons. First, the nontrivial gap comes from different design concepts. Software algorithm designers ignore the design for temporary resources at runtime, such as register wiring and clock delay, which is essential for hardware engineers. Second, FPGAs are notoriously requiring specific technical expertise and significant effort to devise, implement, and validate the system. For many organizations, these skills are either unavailable, or uneconomic as the development cost exceeds potential savings. Oftentimes, people ask more about how to "use" graph accelerators to achieve high performance on their graph algorithms, other than "developing" graph processing projects on FPGA. Therefore, the most challenging issue for high-level application designers is to build an easy-to-use programming environment for graph accelerators. 

	In Figure \ref{fig-apps}, we compare different system development approaches. Building a graph computing system from the scratch is difficult and time-consuming. Using existing graph accelerators/ISA can reduce the difficulty of system implementation, but one still need to spend time on understanding the architecture and developing design experience. High-Level Synthesis (HLS) tools are designed to translate the high-level language like C to HDL (e.g. Verilog) to synthesis on FPGA. However, these general-purpose HLS tools (or translators) often waste much time on compiling. It is highly desirable to simplify the programming procedure. 
	
	On the other hand, domain-specific language (DSL) is popular for application design in certain domain because of the flexibility and easy-to-use design. DSLs for graph processing \cite{graphit,falcon} are proposed to generate fast implementations for graph applications. Accordingly, domain-specific programming system can help to bridge the gap between design and performance. In addition, A newly-designed DSL collaborated with a translator is the most efficient way for agile developers. 
	
	General-purpose translators usually provide inefficient codes which can hardly provide the performance as expected. Although there are compilers\cite{vivado,spatial,chisel} that can translate high-level language such as C and Scala to Verilog used by FPGAs, the generated code lacks efficiency. There are two main reasons. First, the code generated from these general-purpose HLS tools leave some FPGA resources under-utilized. They often use as many registers and logic units as they can to provide temporary calculation results and recycle (in fact, rarely) them after use. Thus, each piece of graph data is considered as a single-register results in resources over-occupation. Second, these translators must keep redundant designs to ensure the correctness and the accommodation of universal application. For example, they concentrate more on syntax analysis and design space exploration so that the intermediate operations are sophisticated and time consuming. Light-weight translator considers hardware design much more than compiler strategies since we only process graph applications.

	We propose JGraph, a FPGA-based graph programming framework, which simplifies the development of high-performance graph processing applications on FPGAs. This system hides the complexity of hardware programming while preserving the accelerator efficiency. To our knowledge, JGraph is the first graph programming system on FPGA.

	Our contributions are listed as follows: 
	\begin{itemize}
	\item{We design an easy-to-program graph DSL consists of programming interface with customized parameters, which improves development efficiency and enables the extensibility of the sophisticated graph applications.} 
	\item {We develop a light-weight efficient translator with communication management and runtime scheduling optimization, which generates efficient hardware codes with the control of data transmission and parallel design.}
	\item{ We propose a FPGA-based graph programming system that integrates the graph DSL and the translator in a user-friendly manner. Our evaluation shows that it maintains high performance and high productivity.}

	\end{itemize}

	\section{Related Work}
	
	There are many studies on FGPA graph processing, including both DSL with general-purpose compilers (or HLS tools) and the hard-to-program graph accelerators without compilers. We compare different design approaches in Figure\ref{language}. Below, we will further explain the current situations and challenges.
	\begin{table}[t]
		\caption{Languages on FPGAs with empirical estimates of PD(Programming Difficulty),TT(Time for Translating), RTL(RTL code performance)}
		\label{language}
		\begin{tabular}{@{}c|ccccc@{}}
			\toprule
			Type                                                                         & Languages     & field & PD     & TT     & RTL    \\ \midrule
			\multirow{3}{*}{HDL}                                                         & Verilog/VHDL  & all   & hard   & short  & high   \\
			& SystemC       & all   & hard   & short  & high   \\
			& OpenCL        & all   & hard   & short  & high   \\ \midrule
			HDL-like                                                                     & Chisel        & all   & middle & middle & poor   \\ \midrule
			\multirow{4}{*}{High-level}                                                  & Vivado HLS    & all   & easy   & middle & poor   \\
			& Spatial       & all   & middle & long   & middle \\
			& GraphIt(C)    & graph & easy   & -      & -      \\
			& Falcon(C)     & graph & easy   & -      & -      \\ \midrule
			\multirow{4}{*}{\begin{tabular}[c]{@{}c@{}}Graph \\ accelerators\end{tabular}} & Graphgen      & graph & -      & short  & high   \\
			& GraVF         & graph & -      & short  & high   \\
			& GraphSoC      & graph & -      & short  & high   \\
			& graphicionado & graph & -      & short  & high   \\ \bottomrule
		\end{tabular}
	\end{table}
	\subsection{General-purpose Language for Graph}
	In general, FPGA programming relies on Hardware Description Languages (HDLs), such as Verilog, VHDL, SystemC, OpenCL, etc. They are designed for hardware developers in almost all fields. Although HDL-based design allows one to achieve high performance, the application programming and FPGA deployment can be time-consuming. Graph accelerators\cite{graphgen,gravf-m,graphops,graphicionado} provide certain kinds of graph interfaces to improve usability. However, they are not flexible enough if we want to implement other graph algorithms other than implemented ones, let alone customized functions. 
	
	High-Level Synthesis(HLS) tools are designed to translate a high-level language to HDL such as Xilinx Vivado HLS\cite{vivado} and Spatial\cite{spatial}, etc. Additionally, HDL-like languages such as  Chisel\cite{chisel} add hardware construction primitives to the Scala programming language to write complex, parameterizable circuit generators. However, the generated code is inefficient and the compiling procedure is time-consuming. 
	
	Furthermore, graph DSL on CPU \cite{graphit} and GPU \cite{falcon} is receiving growing interests. For example, Graphit\cite{graphit} is a C-based DSL generate high performance C project. Falcon\cite{falcon} DSL  is based CUDA and OpenCL on GPU, which can achieve the performance close to that of a handcrafted code. The programming challenge, particularly in hardware configuration, causes more time overhead in detecting failures and error prone. The hardware abstraction and code generation are optimized with explicitly parallelism.
	In summary, each of the languages and frameworks mentioned above has their pros and cons. For graph processing, FPGA programming system is required to gain better performance with friendly and productively programming experience. Our goal is to design an easy-to-program graph DSL that can facilitate hardware development to unleash the full potential of accelerators.

	\subsection{Programming Frameworks on Graph Accelerators}
	There are three categories of graph processing interfaces on FPGAs according to the abstraction granularity. More details of graph processor on FPGA are shown in Table \ref{tab-programmable}.
	\begin{table*}[t]
		\caption{Programmable interface and Instructions for graph processing on FPGA accelerators}
		\label{tab-programmable}
		
		\begin{tabular}{|c|c|c|c|c|c|}
			\hline
			\multicolumn{2}{|c|}{\multirow{2}{*}{\textbf{\begin{tabular}[c]{@{}c@{}}Graph frameworks\\ on FPGA platform\end{tabular}}}}                                                & \multirow{2}{*}{\textbf{\begin{tabular}[c]{@{}c@{}}Algorithm supported\\ (PR/BFS/SSSP)\end{tabular}}} & \multicolumn{3}{c|}{\textbf{Function supported}}                                                                                                                                                                                                                                                             \\ \cline{4-6} 
			\multicolumn{2}{|c|}{}                                                                                                                                                     &                                                                                                                  & \textit{\textbf{\begin{tabular}[c]{@{}c@{}}Basic\\ Processing\end{tabular}}}                           & \textit{\textbf{\begin{tabular}[c]{@{}c@{}}Edge Data\\ Processing\end{tabular}}}      & \textit{\textbf{\begin{tabular}[c]{@{}c@{}}Communication\\ manager\end{tabular}}}                     \\ \hline
			\multirow{7}{*}{\begin{tabular}[c]{@{}c@{}}single\\ FPGA\end{tabular}}                       & GraphGen'14                                                                 & \begin{tabular}[c]{@{}c@{}}app-specific graph\end{tabular}                                                    & update-function(v)                                                                                        & -                                                                                     & -                                                                                                        \\ \cline{2-6} 
			& GraphSoc’15 (multi-PE)                                                          & SpMV/ etc.                                                                                                        & \begin{tabular}[c]{@{}c@{}}SND,RSV,ACCU, UPD,etc.\end{tabular}                                          & \begin{tabular}[c]{@{}c@{}}Receive, sendLC,\\LS, LMSG,etc.\end{tabular}           & DC,NOP,HALT,etc.                                                                                         \\ \cline{2-6} 
			& GraVF’16                                                                    & basic/ etc.                                                                                                       & Apply, Scatter                                                                                             & -                                                                                     & -                                                                                                        \\ \cline{2-6} 
			& Graphicionado’16                                                            & \begin{tabular}[c]{@{}c@{}}Collaborative Filtering /etc.\end{tabular}                                       & \begin{tabular}[c]{@{}c@{}}Reduce(v,r), Apply(v),\\ Process\_Edge\end{tabular}                          & -                                                                                     & -                                                                                                        \\ \cline{2-6} 
			& \begin{tabular}[c]{@{}c@{}}GraphOps’16 (library)\end{tabular}           & \begin{tabular}[c]{@{}c@{}}SpMV/ conduct/ vcover/ etc.\end{tabular}                                              & \begin{tabular}[c]{@{}c@{}}Data\_block\end{tabular}         & \begin{tabular}[c]{@{}c@{}}Control\_block\end{tabular} & \begin{tabular}[c]{@{}c@{}}utility\_block\end{tabular}                          \\ \cline{2-6} 
			& \begin{tabular}[c]{@{}c@{}}FPGP’16 (Graphlab)\end{tabular}          & BFS                                                                                                              & BFS\_kernel                                                                                               & \multicolumn{1}{l|}{Data\_controller}                                                 & Memory controller                                                                                        \\ \cline{2-6} 
			& Hitgraph’19                                                                 & SpMV/ WCC                                                                                                         & \begin{tabular}[c]{@{}c@{}}Apply\_update,\\ Process\_edge\end{tabular}                                    & -                                                                                     & -                                                                                                        \\ \hline
			\multirow{3}{*}{\begin{tabular}[c]{@{}c@{}}FPGA \\ with \\ specific \\ storage\end{tabular}} & \begin{tabular}[c]{@{}c@{}}Graphlet’11 (off-chip) \end{tabular} & \begin{tabular}[c]{@{}c@{}}graph counting\end{tabular}                                                        & Graph Processing Elements                                                                                 & -                                                                                     & \begin{tabular}[c]{@{}c@{}}interconnect network,\\ run-time unit\end{tabular}    \\ \cline{2-6} 
			& \begin{tabular}[c]{@{}c@{}}GraFBoost’18 (flash storage)\end{tabular}      & BC/ etc.                                                                                                          & \begin{tabular}[c]{@{}c@{}}Vertex\_update, \\ finalize, is\_active\end{tabular}                            & Edge\_program                                                                         & -                                                                                                        \\ \cline{2-6} 
			& GPOP’19 (HBM2)                                                                & SpMV/ WCC/ etc.                                                                                                    & \begin{tabular}[c]{@{}c@{}}algorithmic parameters: \\ number of graph,\\ partation, iteration\end{tabular} & -                                                                                     & -                                                                                                        \\ \hline
			\multirow{2}{*}{\begin{tabular}[c]{@{}c@{}}Multiple\\ FPGA\end{tabular}}                     & Foregraph’17                                                                & WCC/ etc.                                                                                                         & Processing elements                                                                                       & Data controller                                                                       & \begin{tabular}[c]{@{}c@{}}Interconnection  controller, \\ Off-chip memory controller\end{tabular} \\ \cline{2-6} 
			& GraVF-M’19                                                                  & WCC/ etc.                                                                                                         & \begin{tabular}[c]{@{}c@{}}vertex kernel:\\ gather, apply, scatter\end{tabular}                             & -                                                                                     & -                                                                                                        \\ \hline
		\end{tabular}
		
	\end{table*}
	
	\subsubsection {Graph IP Cores}
	
It is desirable to build package (IP Cores) dedicated to certain graph algorithms on FPGA accelerators. There are many works \cite{GPOP,fpgp,graphlet,zhou3hitgraph,gravf-m} developing IP cores for several key graph kernels (e.g., BFS kernel) with changeable parameters. We can process certain graph algorithms on them with low flexibility. For example, GPOP\cite{GPOP} develops 4 IP cores for graph kernels: SpMV, PR, SSSP, and WCC kernels to process partitioned graph data. 
	\subsubsection {Graph APIs and Libraries}
	
	Graph APIs \cite{grafboost,foregraph} and libraries\cite{graphops} include functions and interfaces about graph-oriented operations, some of which provide graph data and communication controllers. For example, Gather-Apply-Scatter (GAS) \cite{graphlab} is a widely used popular model for graph. Vertex processing is the core operations in most of these works. Edge data processing reflects the graph data control strategy in kernels. Additionally, on multi-FPGA platform and FPGA with specific memory architecture such as Hybrid Memory Cube (HMC), communication manager is often used for interconnect network control and runtime scheduling. 
	
	\subsubsection {Graph Instructions}
	
	Some works provide a few graph instructions abstracted from graph atomic operations. For example, GraphSoc\cite{GraphSoc} provides a soft processor Instruction Set Architecture (ISA) to implement specific repetitive operations on graph nodes and edges observed in sparse graph computations. Fine-grained abstraction improves the flexibility for programming.
	
	\section{JGraph System Design}
	\begin{figure}[t]
		\centering
		\includegraphics[width=\linewidth]{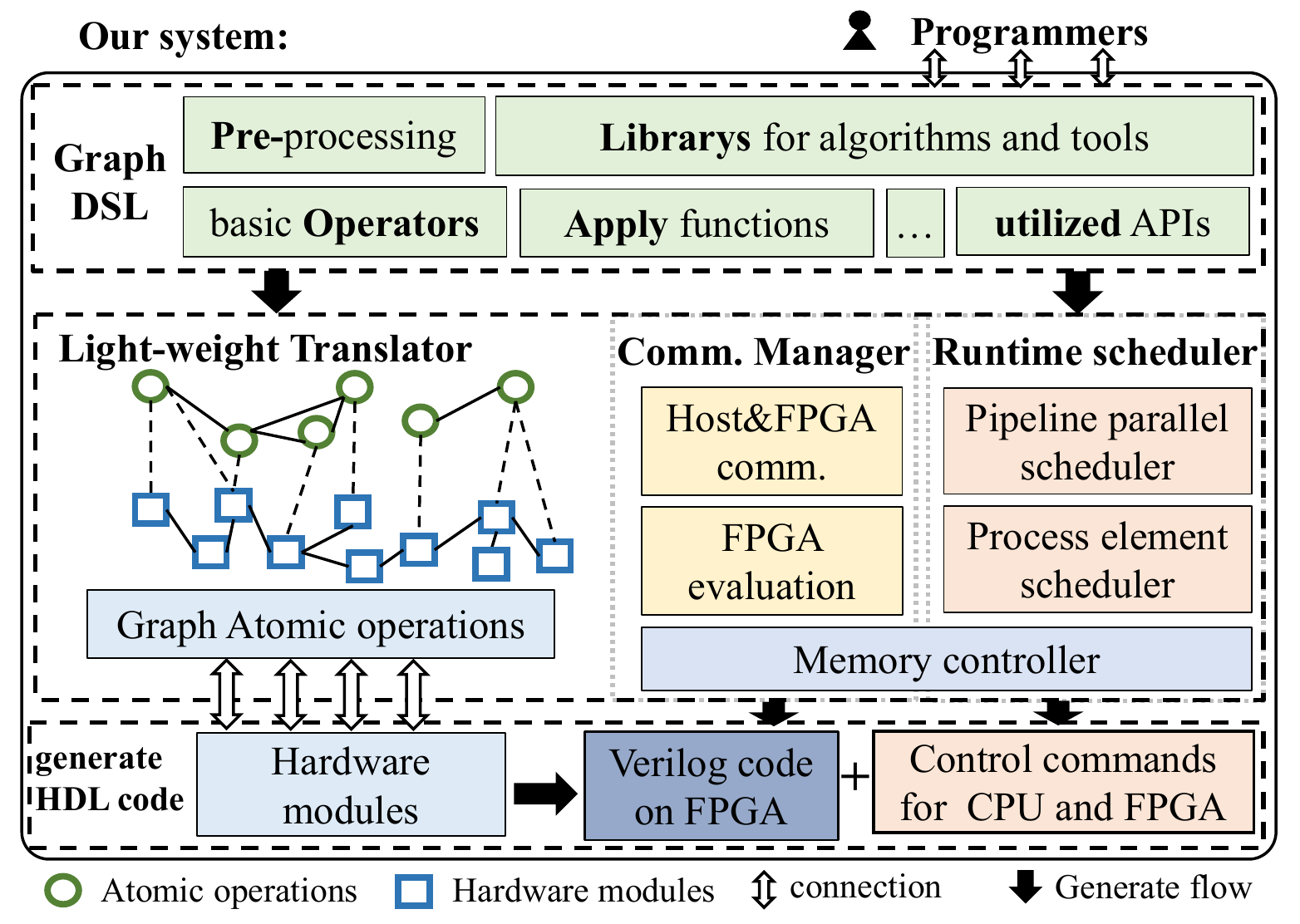}
		\caption{Our graph programming framework JGraph}
		\label{ours}
	\end{figure}
	\
	We design a FPGA programming environment called JGraph. The software system design module diagram is shown in figure \ref{ours}. Our system consists of two main parts, 1) a high-level graph DSL, 2) a light-weight translator with communication manager and runtime scheduler. 
	
	Given the components mentioned above, our system can generate efficient C and HDL code. The C code will be executed on CPU, mainly including data transmission control commands. Meanwhile, the part of (accelerated) HDL code will be deployed on FPGA. In this step, there is a conversion from Chisel HDL to Verilog HDL that can be executed on the FPGA. The above two parts work together to complete the data transmission and graph operations.
	
	The framework supports a variety of graph processing programming models, including the extraction and encapsulation of a variety of classic graph operators. The upper layer provides an easy-to-use programming language and flexible graph functions for users. The framework designs efficient and light-weight translator technology, mapping graph functions to high-performance HDL code of FPGA design modules. In this way, we can program easy-to-use high-level code and complete efficient hardware implementation.

	\section{Graph DSL Design}
	
	Our graph programming language is based on the object-oriented language Scala. We use Chisel\cite{chisel}, a state-of-the-art HDL language with Scala as the intermediate language. The decoupling of graph scheduling and graph algorithm is convenient for translator optimization. Due to the diversity of graph algorithm, the adoption of the GAS programming model can abstract the process of graph algorithms flexibly. In this way, we peel off the graph traversal and data organization from algorithm design.
	
\begin{figure}[t]
		\centering
		\includegraphics[width=\linewidth]{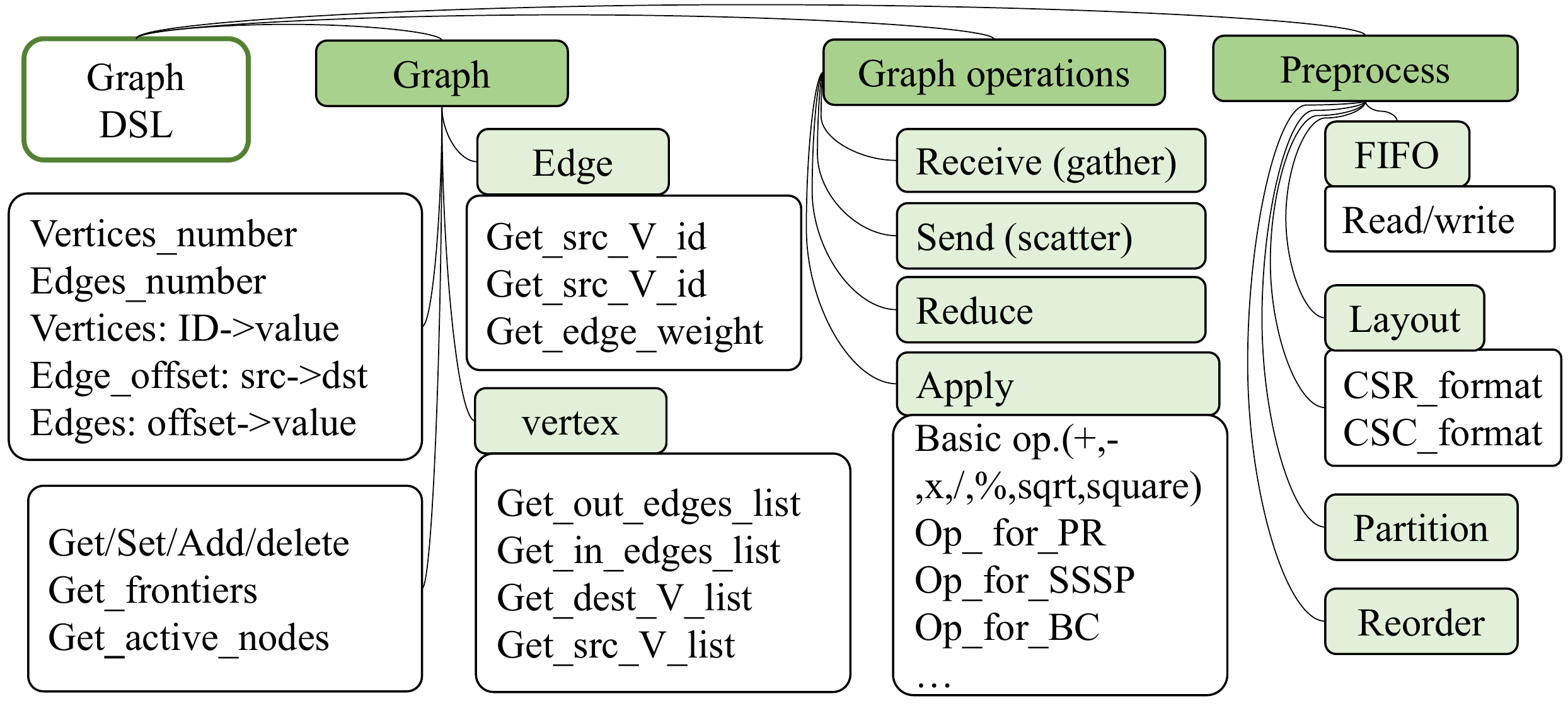}
		\caption{Graph functions that our framework provides}
		\label{dsl}
	\end{figure}

	\textbf{Graph operators abstraction:} Graph operators are defined with the atomic operators for graph processing applications, for example, getting out-edges of certain vertex, getting neighbors of this vertex, and so on. We designed our DSL from the perspective of programmers, accommodating accelerator implementation to user interfaces. There are three main parts in our DSL: \textit{Graph data}, \textit{Graph operations} and \textit{Preprocessing}, which are shown in Figure \ref{dsl}. One can easily understand the logic of our interfaces design and quickly rebuild the custom functions if necessary. In Algorithm 1, we give a pseudocode\ref{alg:BFS} of BFS using our Jgraph as an example.

	\subsection{Functions for \textit{Graph data}} \textit{Graph data} consists of edge sets, vertex sets, and their values. We design three parts for graph data fetching and modifying.  They act as key attributes no matter which formats of graph data we use. We choose Compressed Sparse Row (CSR) data format because CSR saves memory and is easy for memory accessing.
	
	\subsubsection{Graph Data}
	The graph data can be represented by three arrays as shown in Figure \ref{dsl}.	We provide interfaces to get and update the graph data value mentioned below. For graph algorithms, for frontiers are often used like a queue including vertices to be processed on this iteration. Additionally, active and inactive nodes are used for partial traversal at certain situations.
	
	\begin{itemize}
		\item{ \textbf{\textit{Vertices}}:} The array index is vertex id and the array value is vertex value.
		
		\item{\textbf{\textit{Edge\_offset}}:} The indexes of this array are source vertex ids and the corresponding values are the offset values of the destination vertex ids.
		
		\item{\textbf{\textit{Edges}}:} The indexes of this array are edge offsets from \textit{Edge\_offset} and the corresponding array values are edge weights. 
	\end{itemize}

	\subsubsection{Graph Vertex}	
	Vertex operations concentrate on obtaining neighbors' information to help traversal.
	\begin{itemize}	
		\item{\textbf{\textit{Update\_Vertex}}:} Updating the vertex value is the most regular operation for data updating. To improve performance and memory access efficiency, the vertex value are often transfered to BRAM in advance.
		\item{ \textbf{\textit{Get\_out\_edges\_list/Get\_in\_edges\_list}}:} We get the out/in edges of certain vertex and return the list of edge id and weight.
		\item{\textbf{\textit{Get\_dest\_V\_list/Get\_src\_V\_list}}:} Through the out/in edges of the vertex we can get the lists of out/in neighbor id.
	\end{itemize}	 
	
	\subsubsection {Graph Edge} 
	Edge operations is similar with vertex operations. We can get and update edge weights.
	\begin{itemize}	
		\item{\textbf{\textit{Get\_src\_V\_id/Get\_dest\_V\_id}}:} We can obtain the source/destination vertices of certain edge with edge id and weight.
		\item{\textbf{\textit{Get\_edge\_V\_weight}}:} Through the id of the edge we can get the weight of this edge.
		
	\end{itemize}

	\begin{algorithm}[t]
		\caption{A Pseudocode of BFS using FAgraph}
		\label{alg:BFS}
		\textbf{Input}: Vertices, Edge\_offset, Edges \\
		\textbf{Output}: Updated vetices 
		\begin{algorithmic}[1] 
			\STATE Graph = \textcolor{magenta}{Read}(graphFile);          \hfill//\textit{FIFO}
			\STATE GraphCSC = \textcolor{magenta}{Layout}(Graph, CSC);     \hfill//\textit{Layout}
			
			\STATE \textcolor{cyan}{Get\_FPGA\_Message}(); \hfill//\textit{comm. manager}
			\STATE \textcolor{cyan}{Transport}(CPU\_ip, FPGA\_ip, GraphCSC);
			\STATE Set \textcolor{blue}{Pipeline} = 8, \textcolor{blue}{PE} = 1;  \hfill//\textit{runtime scheduler}
			
			\STATE get Graph with Vertices, Edge\_offset, Edges 
			\STATE \textcolor{magenta}{Reorder}(GraphCSC)   \hfill//\textit{Reorder (optional)}
			\STATE \textcolor{magenta}{Partition}(GraphCSC) \hfill//\textit{Partition (optional)}
			
			\WHILE{\textcolor[rgb]{0,0.8,0}{Get\_active\_vertex}()}      
			\STATE v = \textcolor[rgb]{0,0.8,0}{Get\_vertex}(i);       \hfill//\textit{Vertex}
			\STATE off = \textcolor[rgb]{0,0.8,0}{Get\_edge\_offset}(v);  \hfill//\textit{Edge\_offset} 
			\FOR {each off}
			\STATE e = \textcolor[rgb]{0,0.8,0}{Get\_edge}(j); \hfill//\textit{Edge}
			\STATE u = \textcolor[rgb]{0,0.8,0}{Get\_src\_V\_id}(e); \hfill//\textit{Receive}
			\STATE tmp = \textcolor[rgb]{0.1,0.5,0}{Apply}(v, e, u);  \hfill//\textit{Apply}
			\STATE updated = \textcolor[rgb]{0.1,0.5,0}{Reduce}(tem1,tem2,...); \hfill//\textit{Reduce}
			\ENDFOR
			\IF {updated != v}
			\STATE \textcolor[rgb]{0,0.8,0}{Set\_Vertex\_value}(updated);
			\ENDIF
			\STATE \textcolor[rgb]{0,0.8,0}{Update\_vertex}(v);
			\ENDWHILE
			
		\end{algorithmic}
	\end{algorithm}

	\subsection{Functions for \textit{Graph Operation}} 
	Graph operation functions are critical functions for users to calculate and update the vertex while traversal.  In the GAS model, we often think like a vertex. The traversal and topology of a graph can be decoupled to messages transferring between vertices. By the way, \textit{Send} and \textit{Receive} are the contract ways and can often be replaced by each other. 
	\begin{itemize}
		\item{\textbf{\textit{Receive}}:} In \textit{Receive}, this vertex receives data from neighbors. The input parameters include source vertex list and data location. This function returns the desired data to compute the new value of the vertex. 
		\item{\textbf{\textit{Reduce}}:} In some case, there are multiple update messages for the certain vertex. We should reduce these message with accumulator \cite{hust2yao} to combine the received messages.
		\item{\textbf{\textit{Send}}:} \textit{Send} acts in the opposite way, sending the updated messages out. The input parameters include destination vertex list and updated data. This function sends the updated data and source messages to neighbors. 
		\item{\textbf{\textit{Apply}}:} After receiving the messages, the \textit{Apply} is used for calculating the new value of this items. Take BFS as an example, the Apply function is the current value plus one after traversal. The basic operators are included such as $+,-,*,/,\%,sqrt,sqare $, etc.\textit{ Apply} contains these operators to be choosed. There are algorithm-aware operators designed for specific graph algorithms. The algorithms are commonly used for graph analysis. So we give the templates for these operators, which can be used conveniently. One can program almost all the graph algorithms through changing the \textit{Apply} interface.
	\end{itemize}
	\subsection{Functions for\textit{ Preprossing}} 
	For most graph algorithms, the graph data needs preprocessing to optimize the execution efficiency.
	\subsubsection{\textbf{FIFO}} FIFO contains file operations, including reading input files, writing data to output files. For graph data in graph database management system such as Neo4j\cite{neo4j}, we can read data from database directly. This step prepares for the data layout conversion. 
	\subsubsection{\textbf{Layout}} There are various graph data layouts, such as CSR, Compressed Sparse Column (CSC), Adjacency matrix,linked list, etc. We provide several functions for data structure transmission. For example, the original data layout Edge list with source and destination vertex, and we can change array list to CSR format. 
	\subsubsection{\textbf{Partition}} For large graph processing, partition methods are various with optimizations. We provide several partition strategies proposed by start-of-art works \cite{powerlyra,pathgraph}. The basic partition is to divide graph into several parts without optimization. We can also separate graph with graph algorithms, such as graph coloring and community detection.
	\subsubsection{\textbf{Reorder}} Graph reordering can improve efficiency of processing and reduce random memory accessing. We provide several reorder methods used in start-of-the-art works\cite{reorder}. We can sort nodes in descending order by degree because higher degree nodes will be accessed more often. We can also use DFS to find several closed neighbors for the certain node.

	\subsection{Hierarchical Interfaces and Libraries} 
	Our flexible and three-level programming libraries implementation  includes atomic operator(instruction) layer, function layer, and algorithm layer considering the three-level abstraction granularity.  Libraries consists of various ordered combinations of graph atomic operations and functions. In our design, the three-level encapsulation is connected to the algorithm library, object and function in the DSL.the three-level libraries are list as follows. 1) The coarse-grained encapsulation is also called the algorithm level, providing algorithm functions with parameters, such as BFS (graph, input, pipelineNum, etc.). 2) The slightly fine-grained encapsulation is the basic operations of graph algorithm, composed of graph functions and the combinations of these functions mentioned above. 3) The fine-grained encapsulation includes sets of exist graph instructions, atimic operations and control commands, such as load\_Vertices, get\_address,etc.).
	
	\section{Light-weight Translator Design}
	Our light-weight translator is tailored to FPGA hardware executions. We design our translator corresponding to the accelerator design, which means the translator corresponds graph operations above to modules on accelerators. Our light-weight design eliminates the complex grammatical and semantic analysis in most compilers, simplifying the design space exploration. The advantage is  efficient build on top of sophisticated state-of-art graph accelerators on FPGA. There is a trade-off between programming flexibility and hardware design complexity. We choose to trade off  general compiling capabilities (most of which are not profitable in certain area) in exchange for much higher performance.

	\subsection{Execution Module on Accelerator}
	
	\begin{figure}[t]
		\centering
		\includegraphics[width=\linewidth]{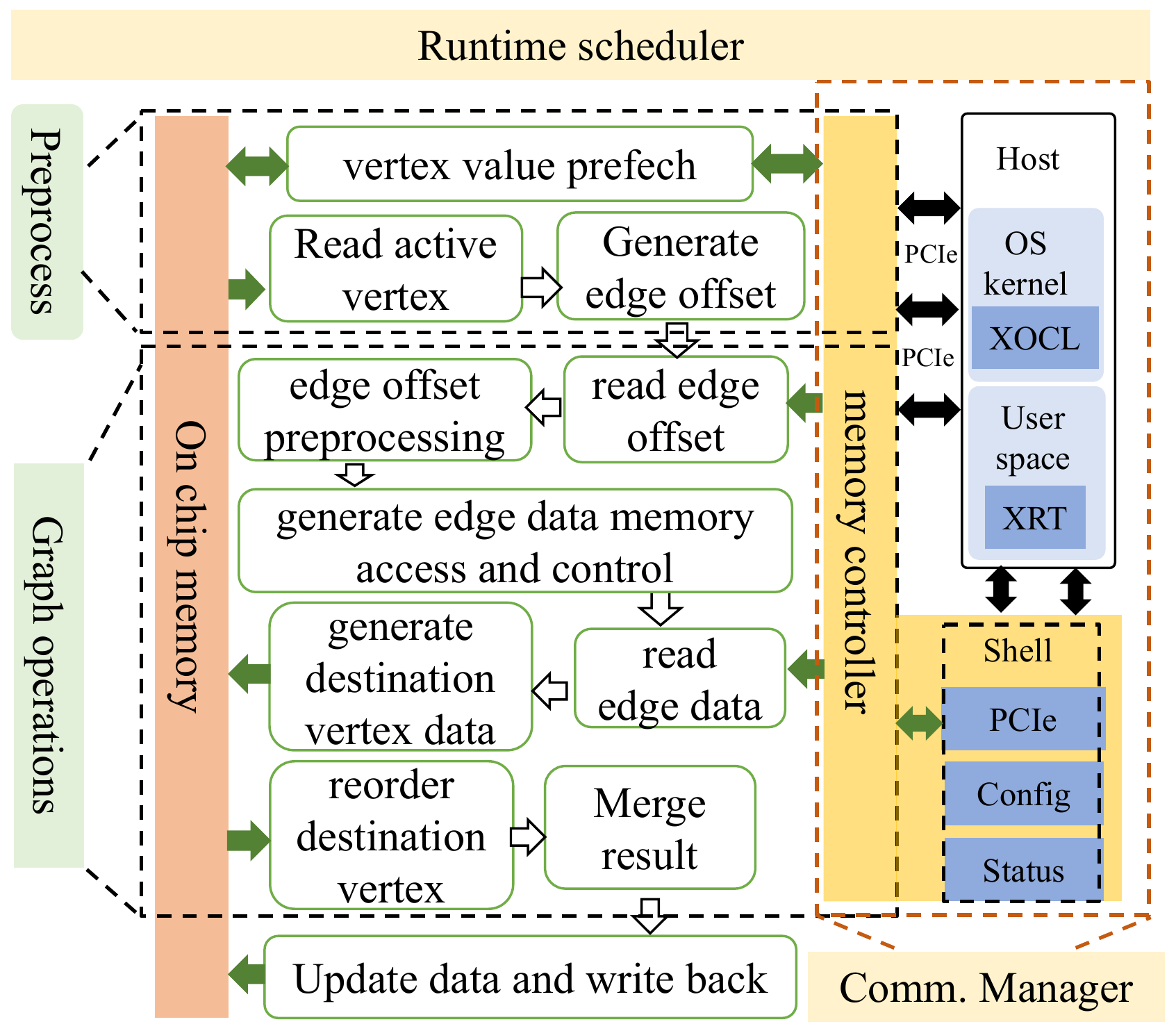}
		\caption{HDL framework on FPGA}
		\label{hdl}
	\end{figure}
	
	In terms of high performance code generation, the CPU and FPGA execute code blocks separately. To make it clearer and easier to debug, C code is executed on the host and the RTL code is executed on the FPGA card. At the same time, the information interaction between the FPGA and the host it can be flexibly grasped. In order to cooperate with the seperated code allocation, the input and updated data executing on FPGA is implicitly transmitted. 
	
	As Figure \ref{hdl} shows, the accelerator module execution flow and the memory access steps are expressed using high-level functions. For GAS model of graph analysis, we can only change several modules to implement different graph algorithms. The execution pipeline are optionally mapped with graph functions provided for developers. Through memory controller provided by FPGA platform Shell, the translator holds the parallel optimization of data transmission, so as to improve the efficiency and reduce the programming difficulty.
	
	\subsection{HLS Adaption and Translator Design}
	
	\textbf{Efficient HLS Implementation and Adaption:} Our designed interfaces for graph processing are closely connected to the hardware modules of graph accelerator in the pipeline execution of graph accelerator. Our graph HLS directly specifies the optimized parallel graph data access operation with parallelism on vertices updating. According to the GAS programming model, the operation steps of the main diagram correspond to the hardware module concisely and efficiently.
	
	\textbf{Translation Design and Optimization:} Different from software algorithm programming, hardware programming often focuses on the design of logic devices.  the existing universal HLS tools translate variable items directly into usable hardware logic units, therefore result in a lot of register applying repeatedly and wastes of logic resources for irregular graph access. In addition, the loop iterations are often automatically transformed into a serious of repeated Arithmetic Logical Units (ALUs) for each iteration. In order to solve these problems and save LUT, we use pipeline stream to improve resource reuse. Aditionally, we focus on the decoupling of data and logic operation to save on-chip memory. We map functions with hardware modules correspondingly and can avoid the redundant computing resources.
	
	\subsection{Management Optimizations}
	To better control the data transfer and instruction configuration, the communication manager and Running manager are designed. CPU sends and receives control instructions and application data through PCIe. Along with the exist memory controller for FPGA board, we can manage the communication between host and accelerator and the running status with parallel scheduler.
	\subsubsection{Communication manager}
	The communication manager between CPU and FPGA board is designed for data transfering and configuration management. In practice, we often ue control shell on host to control the FPGA board. The control shell for host consists of OS kernel controller XOCL and user space controller Xilinx Runtime (XRT). XRT is low-level communication layer including APIs and drivers between the host and the FPGA board while Deployment Target Platform (also called DSA) is the communication layer that is physically implemented and flashed into the board. We can get FPGA running status andsend control instructions through these tools. We also abstract several easy-to-use interfaces to help status transfer and configuration management.
	
	\subsubsection{Runtime scheduler}
	The runtime scheduler consists of two parts: the parallel pipelines scheduling and processing elements (PEs) scheduling, aiming at parallelism management for the whole project. Pipeline deployment is the common way for FPGA to implement parallel computing and accelerate iterations. Each PE is a single processor so that we can deploy several PEs on FPGA to utilize logic cells. In practice, the degree of parallelism for FPGA applications usually depend on the number of pipelines and the processing elements \cite{zhou3hitgraph}. We can specify a specific number of pipelines and PE for the program to achieve flexible parallelism.

	\section{Evaluation}
	We conduct experiments using the Xilinx Alveo U200 Data Center accelerator A-U200-A64G-PQ-G FPGA cards with PCI Express Gen3x16 compliant cards. The target FPGA device has 1,182K LUTs, 2,364K  registers, 6,840 slice DSPs, 960 UltraRAMs and 64 GB DDR4 DRAM. We synthesize, place-and-route, and simulate our designs using Xilinx Vivado Design Suite 2019.3 and SDAccel 2019.3\cite{vivado}. We use XRT, xbutil to flash and configure FPGA card. 
	The real-life graphs are obtained from the Stanford network dataset repository\cite{snap}. We take BFS as the graph algorithm example. The throughput refers to the number of Traversed Edges Per Second (TEPS), computed as the total number of traversed edges divided by the execution time.
	
	\subsection{Application Extensibility}
	To generally show the extensibility of our programming system, we compare several graph accelerators and frameworks that provide programming interfaces or instructions for graph processing. We also use an intuitive method to reflect the capicity and extensibility of our design, as we can see in 
	Figure \ref{tab-exten}.  We provided more than 25 interfaces for people to flexibly program graph algorithms, much more than the available functions and parameters provided in recent works. This also implicitly shows that we have more categories of programmable graph algorithms.
	
	\begin{table}[t]
		\centering
		\caption{Comparations of graph atomic operators with accelerators and programming environment}
		\begin{tabular}{@{}ccl@{}}
			\toprule
			\textbf{Accelerators}  & \textbf{Num} & \multicolumn{1}{c}{\textbf{Graph atomic operators}}                                                                                                                              \\ \midrule
			\textbf{GraFBoost’18} & 4            & \begin{tabular}[c]{@{}l@{}}edge program, vertex update, \\ finalize,and is active.\end{tabular}                                                                                  \\ \midrule
			\textbf{Foregraph’17}           & 5            & \begin{tabular}[c]{@{}l@{}}interconnection controller, \\ off-chip memory controller, \\ a data controller, \\ a dispatcher, \\ several processing\\ elements (PEs).\end{tabular} \\ \midrule
			\textbf{GraphOps’16}           & 7            & \begin{tabular}[c]{@{}l@{}}ForAllPropRdr,NbrPropRed, \\ ElemUpdate,QRdrPktCntSM,\\ UpdQueueSM,EndSignal,MemUnit\end{tabular}                                                     \\ \midrule
			\textbf{GraphSoc’15}           & 17           & \begin{tabular}[c]{@{}l@{}}SND, RCV,ACCU,UPD,SAR,\\ DC,B,BNZ,NOP,HALT,LC,LS,\\ LMSG,DC+SND,DC+LS+LMSG\end{tabular}                                                               \\ \midrule
			\textbf{FAgraph}                   & \textbf{25+}          & Details shown as figure \ref{dsl} in section IV                                                                                                                               \\ \bottomrule
		\end{tabular}
		\label{tab-exten}
	\end{table}
	\subsection{Accelerator Code Efficiency}
	 Tailored to the architecture of FPGA, our system can generate high-performance hardware code with less code lines, compared with general-purpose FPGA translators such as Spatial\cite{spatial}, Vivado\cite{vivadohls}, Chisel\cite{chisel}, etc. As Table \ref{tab-performance} shows, we can process BFS traversal within tens of seconds (up to 300 MTEPS). The data communication and scheduling control at runtime is critical especially when the hardware resources are limited. Table \ref{tab-performance} gives the number of hardware code lines and throughput (TP), representing the generated code efficiency and graph data processing capability, respectively. The running time (RT) includes the compilation time, the data preprocessing time and the algorithm execution time.
	
	\begin{table}[]
		\centering
		\caption{ Status for the generated code efficiency and graph data processing capability }
		\label{tab-performance}
		\begin{tabular}{@{}cc|cc|cc@{}}
			\toprule
			\multirow{2}{*}{\textbf{Works}}                        & \multirow{2}{*}{Code lines} & \multicolumn{2}{c|}{\begin{tabular}[c]{@{}c@{}}email-Eu-core \\ 1,005 vertices \\ 25,571 edges\end{tabular}} & \multicolumn{2}{c}{\begin{tabular}[c]{@{}c@{}}soc-Slashdot0922\\  82,168 vertices \\ 948,464 edges\end{tabular}} \\ \cmidrule(l){3-6} 
			&                                      & RT(s)                     & \begin{tabular}[c]{@{}c@{}}TP\\ (MTEPS)\end{tabular}                     & RT(s)                   & \begin{tabular}[c]{@{}c@{}}TP\\ (MTEPS)\end{tabular}                  \\ \midrule
			Spatial                                                & 128                                  & 11.8                                & 19.53                                                                            & 29.3                              & 28.02                                                                         \\
			\begin{tabular}[c]{@{}c@{}}Vivado\\   HLS\end{tabular} & 54                                   & 12.6                                & 199.34                                                                           & 33.8                              & 205.88                                                                        \\
			FAgraph                                                & 35                                   & 5.3                                 & 314.72                                                                           & 15.1                              & 409.04                                                                        \\ \bottomrule
		\end{tabular}
	\end{table}
	
	\begin{figure}[h]
		\centering
		\includegraphics[width=\linewidth]{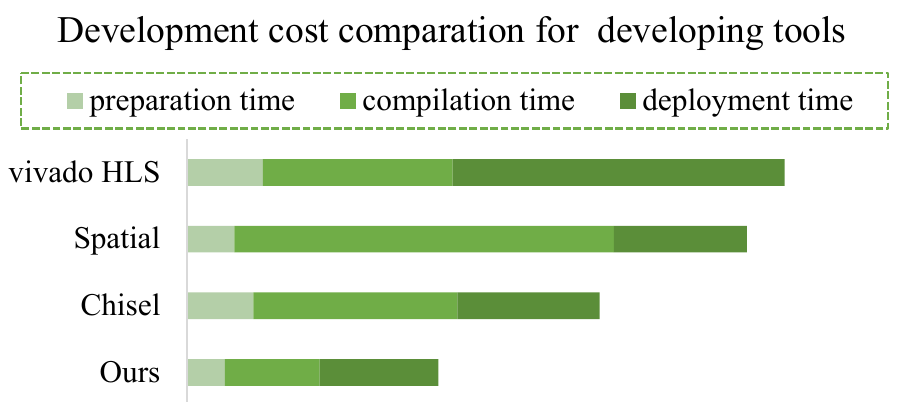}
		\caption{Three periods for programming on FPGA}
		\label{cost}
	\end{figure}
	To show the comparation of development time cost intuitively, Figure \ref{cost} is an example for the detailed stages of running time, representing the development costs on average. The critical steps include program preparation time, system compilation time and environment deployment time. Apart from running and deploying time, development costs  also contain variable time of resources cost,  variable time of manpower, etc. We can see from the figure that our design can reduce the development time and the compilation cost, while maintaining good performance.

	\section{Conclusion}
	
	We propose a graph programming system on FPGA named JGraph. One can program graph algorithms with the functions in the graph DSL, extending the exist graph algorithms. The light weight translator generates efficient code with the control of data transmission and parallel flexibility. To the best of our knowledge, our work is the first graph programming system with DSL and translator on FPGA platform and can generate efficient code within tens of seconds. The design can help to decrease the difficulty of graph programming on FPGA accelerators.

	
	\bibliographystyle{IEEEtran}
	\bibliography{IEEEexample}

\end{document}